\documentclass[pre,amsmath,twocolumn, amssymb,notitlepage]{revtex4}

\usepackage{dcolumn}
\usepackage{lipsum}
\usepackage{graphicx}
\usepackage{placeins}
\usepackage{float}
\usepackage{bm}
\usepackage[T1]{fontenc}
\usepackage{color}
\usepackage{url}
\usepackage[bf]{subfigure}
\usepackage{rotating}
\usepackage{amsfonts}
\usepackage{multirow}	
\usepackage{xcolor}
\usepackage[utf8]{inputenc}

\begin{document}

\title{Topology and Dynamics in Complex Networks: The Role of Edge Reciprocity}

\author{Paulo J. P. de Souza$^1$}
\author{Cesar H. Comin$^2$}
\author{Luciano da F. Costa$^1$}

\affiliation{$^1$S\~ao Carlos Institute of Physics, University of S\~ao Paulo, S\~ao Carlos, SP, Brazil\\}
\affiliation{$^2$Department of Computer Science, Federal University of S\~ao Carlos, S\~ao Carlos, SP, Brazil\\}

\begin{abstract}
A key issue in complex systems regards the relationship between topology and dynamics. In this work, we use a recently introduced network property known as steering coefficient as a means to approach this issue with respect to different directed complex network systems under varying dynamics. Theoretical and real-world networks are considered, and the influences of reciprocity and average degree on the steering coefficient are quantified. A number of interesting results are reported that can assist the design of complex systems exhibiting larger or smaller relationships between topology and dynamics.
\end{abstract}

\maketitle

\section{Introduction}
A particularly interesting question regarding complex systems is to which extent their topology influences the respective dynamics.  In the case of complex systems represented in terms of networks, this question can be approached by considering relationships between topological and dynamical properties. A primary candidate for topological measurement is the degree, which is known to strongly influence both the topology and the dynamics of complex networks~\cite{da2007correlations, comin2014random, adamic2001search, gomez2008entropy}. Similarly, the activation of nodes during a diffusive dynamics represents a natural way to study the dynamics in networks.  The relationship between the degree and the activation at each node therefore provides a reference way to approach the relationship between topology and dynamics of complex systems. Given its importance, this relationship has been called~\cite{de2017network} \emph{steering coefficient}.  

Previous studies have identified maximum possible steering coefficient for undirected networks~\cite{norris1998markov} and networks for which the in-degree is equal to the out-degree for every node~\cite{da2007correlations}.  More recently~\cite{comin2014random,lambiotte2014random}, this question was approached with respect to modular networks, and the results show that different steering coefficients can be obtained for distinct communities in the same network.  Many other important questions regarding the steering coefficient remains unanswered, including the effect of having the reciprocity of a network influenced by measurements such as degree, matching index, clustering coefficient.  This is a particularly important problem because the most sophisticate behaviors of the steering coefficient take place for directed networks, which entail the question of reciprocity.   The current work reports such an investigation, with respect to theoretical and real-world networks.  

More specifically, we consider traditional random walks in four theoretical complex network models, as well as four real-world networks, with varying reciprocity.  We use the degree, clustering coefficient and matching index as a way to assert the edge reciprocity. We start with a network having reciprocity 1. Then, edges are rewired until a given reciprocity is attained. Edges are selected according to a probability that is proportional to the aforementioned properties and connected to new nodes, which are uniformly selected. A random walk dynamics is then applied to the generated network, and the node activations are recorded in order to obtain the respective steering coefficients.

Several interesting results have been obtained, including the fact that the relationship between the frequency of visits and networks degree, as expressed by the steering coefficient, is less affected by rewiring in networks with higher average degree or small diameter.  Also, we have that, for the Watts-Strogatz model, the steering coefficient exhibited a valley as the reciprocity decreases, with the minimal value corresponding to the situation where the random walk dynamics will be less affected by the topology.

This article is organized as follows.  First, we present the basic concepts and methods, including the adopted models, reciprocity, the steering coefficient, and the rewiring procedure.  The experiments and results are then presented and discussed, and the main results and prospects for future investigations are provided in the concluding section.

\section{Material and Methods}
The topological theoretical networks most frequently used to represent natural systems include the Barab\'asi-Albert (BA) model~\cite{BA}, which is characterized by a power law degree distribution. Another well-studied model corresponds to Watts-Strogatz (WS)~\cite{WS} networks, consisting of a regular network such as a ring or lattice with high clustering coefficient, randomly rewired so as to have edges connecting different regions.  We also have the uniformaly random model of Erd\H{o}s-R\'enyi (ER)~\cite{ER}, which is defined by a fixed connection probability between each pair of nodes. Finally, we have the Waxman (WX)~\cite{WX} networks, which are an extension of the ER model for an exponential probability given by: 

\begin{equation}\label{eq:waxman}
\rho_{ij} = e^{-\alpha d_{ij}}
\end{equation}

where $d_{ij}$ is the euclidean distance between nodes $i$ and $j$ and $\alpha$ is a parameter that controls the network average degree.

The WS model was built so that the clustering coefficient is high but the average shortest path length is low, ensuring a small diameter. In the WX model were modeled with the exponential parameter that creates the desired properties of high clustering and large diameter. 

In addition, we used real-world undirected networks respective to airport routes~\cite{openflights}, urban networks and co-authors in high-energy publishing~\cite{hep_papers}. In the airport network, the edges represent the routes and the nodes stand for the airports.  In the cities networks, the nodes are the street crossings or end-lines, while the edges are associated to the streets. This network has a narrow degree distribution, and large average short paths and clustering coefficient. In the co-author network, the nodes are the authors and the edges represent if they have already worked together. Compared to the airport network, the joint authorship network has a relatively large average shortest path and clustering coefficient.

\subsection{Topological properties}
In order to quantify the directionality of the considered networks, we used the reciprocity index, defined~\cite{wasserman1994social} as: 
\begin{equation}
r = \frac{E^\leftrightarrow}{E^\rightarrow + E^\leftrightarrow}
\end{equation} 
where $E^\leftrightarrow$ represents the number of bidirectional edges, and $E^\rightarrow$ is the quantity of directed edges. The matching index~\cite{costa2011analyzing} of an undirected edge reflects the connectivity between the neighbors of the nodes that are respectively connected. The matching index of an edge that connects the node $i$ with the node $j$ can be calculated by:
\begin{equation}
\mu_{ij} = \frac{\sum_{k\neq i,j} a_{ik}a_{jk}}{\sum_{k\neq j}a_{ik} + \sum_{k\neq i}a_{jk}}
\end{equation}

\subsection{Steering Coefficient}
Random Walk dynamics can be interpreted as a diffusion process related to an agent moving in a network and activating its nodes as it visits them. Random walks are extensively studied in Physics, with potential for modeling several real-world systems, as well as applications including the Page Rank algorithm~\cite{xing2004weighted}.  One of the main current problems in network science involves relating topology and dynamics~\cite{ben2000diffusion,van1992stochastic}, such as predicting the activation of nodes given certain topological properties~\cite{da2007correlations, comin2014random}.

\begin{figure}[ht]
	\begin{center}
        \includegraphics[scale=0.45]{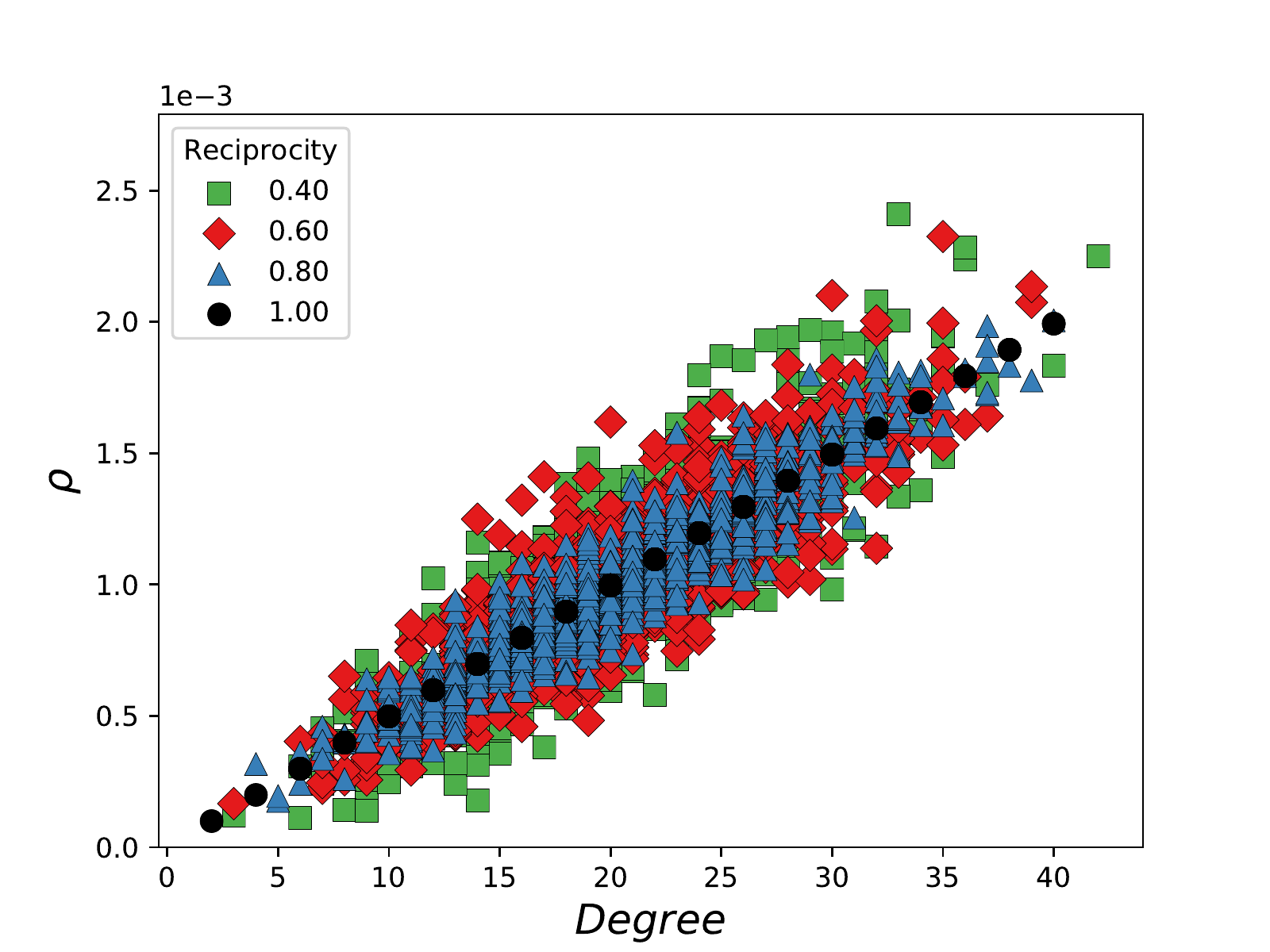}
		\caption{The scatterplot showing the relationship between the node activation and degree for an Erd\H{o}s-R\'enyi model with $1.10^3$ nodes and $<k>\approx10$.  Interestingly, the correlation is steadily lost as the directionality increases. }
		\label{fig:Sterring_coeff}
	\end{center}
\end{figure}

In the case where the in-degree is identical to the out-degree, the node activation becomes fully correlated with the respective degree.  Observe that undirected networks, by obeying this condition, also exhibit a fully correlated relationship.  However, such a relationship is not well understood when this property is not satisfied. Figure~\ref{fig:Sterring_coeff} illustrates how the full correlation between node activation and degree is progressively lost as the reciprocity decreases. 

One direct way to quantify the relationship between topology and dynamics consists in calculating the Pearson correlation coefficient between two respective measurements.  In this work, we will use the Pearson correlation coefficient of the relationship between the node activation and degree, henceforth called \emph{steering coefficient}~\cite{de2017network}.  Such a name suggests the effect of the network topology in driving its dynamics.

In order to investigate how the relationship between the degree and dynamics change for different levels of reciprocity, it is necessary to rewire the considered networks so as to change their reciprocity.  In this work, the rewiring is performed as follows.  First, we choose a node $i$I with probability preferential to specific measurements (node degree and clustering coefficient), and then choose among the neighbors of $i$, with an analogos preferential criterion, another node $j$, defining the directed link from $i$ to $j$.  The tip of this edge is then randomly assigned to another node with uniform probability.  This basic step is repeated until the sought reciprocity level is achieved.  In the case of the matching index, the edge is chosen directly with probability proportional to the matching index, and a similar reassignment is implemented.

So as to avoid breaking the network into two or more disconnected components, just the largest strong connected component is kept.  Observe that the network average degree remains constant in the rewiring process.  As a consequence of rewiring, cycles are broken and, consequently, the clustering coefficient and the matching index decreases.

\section{Results and Discussion}

The study of the relationship between degree and dynamics for varying reciprocity levels was performed with respect to several theoretical network models, as well as some real-world structures.

\begin{table}[ht]
	\begin{center}
    \begin{tabular}{c|l*{7}{c}|r}
		Network  & N & $\overline{k}$ & $\overline{cc}$ & $\overline{\mu}$ & $\overline{l}$\\
        \hline
        BA & 3000 & 6.0 & 0.01 & 0.023 & 4.0\\
        ER & 3000 & 6.0 & 0.004 & 0.002 & 3.5\\
        WS & 3000 & 6.0 & 0.585 & 0.292 & 22.58\\
        WX & 3000 & 6.0 & 0.148 & 0.692 & 11.0\\
        \hline
        Formosa & 3588 & 3.40 & 0.030 & 0.013 & 34.40\\
        Darwin & 2753 & 2.30 & 0.109 & 0.047 & 33.00\\
        Airports & 3010 & 11.60 & 0.602 & 0.123 & 4.0 \\
        Collabo. & 5835 & 4.70 & 0.604 & 0.168 & 7.0
    \end{tabular}
    \caption{Topological properties of the networks used in the analysis. N is the number of nodes, $\overline{k}$ is the average in-degree, $\overline{cc}$ the average clustering coefficient, $\overline{mi}$ the average matching index and $\overline{l}$ the average shortest path length. We note that the WS model has a rewiring parameter of $\rho = 5\times 10^{-3}$.}
    \label{tab:properties}
	\end{center}
\end{table}

\begin{figure}[ht]
	\begin{center}
    	\includegraphics[scale=0.55]{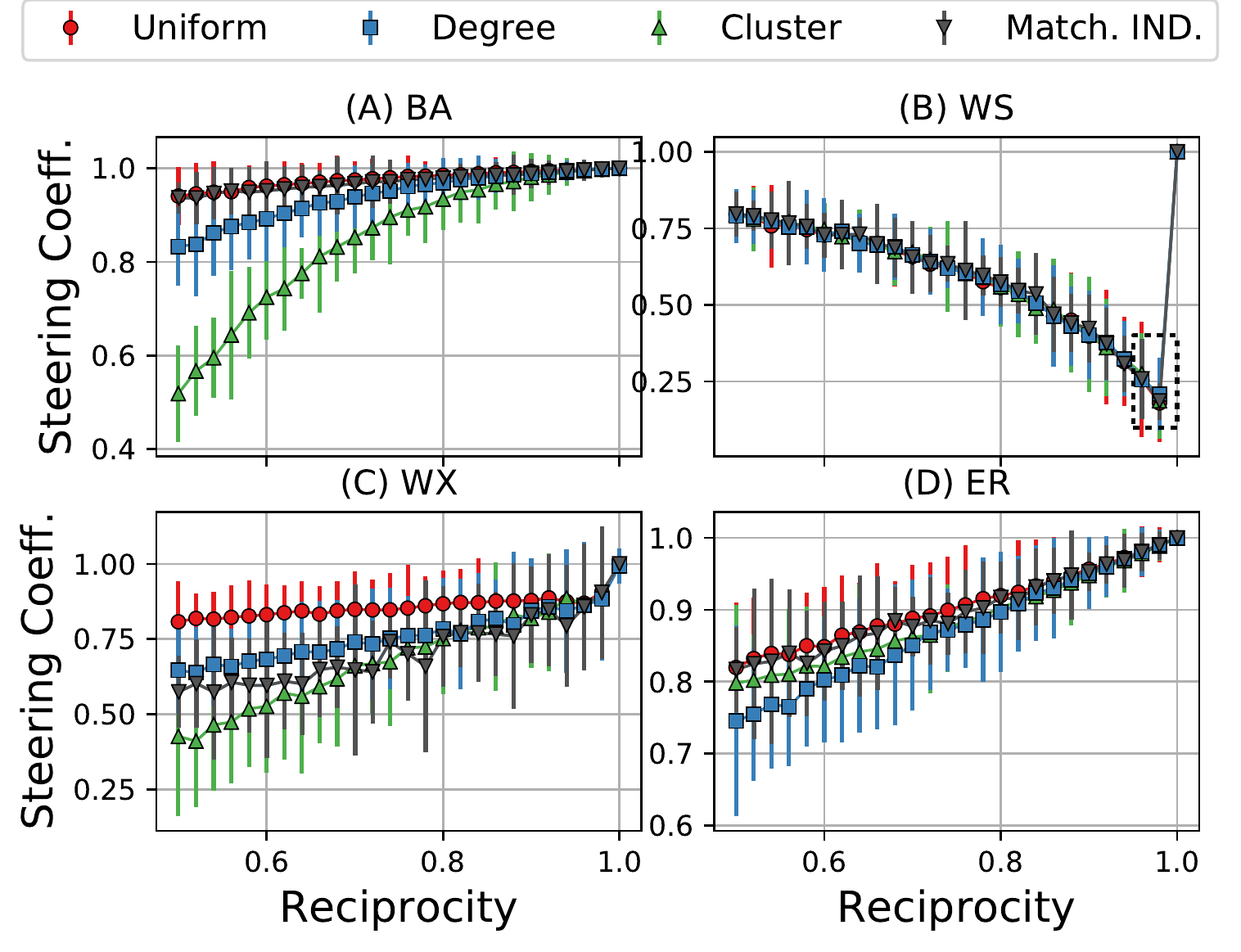}
		\caption{Steering coefficient as a function of edge reciprocity when considering an uniform rewiring strategy as well as rewiring probabilities proportional to the degree, the clustering coefficient and the matching index.}
	    \label{fig:theoretical}
	\end{center}
\end{figure}

We start by discussing the results obtained for four theoretical networks models, namely BA, WS, WX and ER. The parameters of the networks used in the simulation are summarized in table~\ref{tab:properties}.
Fig.~\ref{fig:theoretical} shows the steering coefficient values in terms of the reciprocity for these models, with respect to the four distinct rewiring strategies (shown in different colors). For the BA networks, we obtained almost constant curves with negative concavity, with the exception of the clustering coefficient strategy. Larger variation of the steering coefficient was observed for the WS and WX networks.  In the former case, the curves respective to the several measurements resulted almost identical, indicating that the type of rewiring strategy did not have a large influence. The uniformly random model ER, showed a small variation of the steering coefficient and linear behavior for all rewire strategies. 

More specifically, in the case of the BA networks, almost identical curves are obtained for uniform and matching index-based rewirings. This is a direct consequence of the fact that the matching index tends to zero at all edges as the reciprocity decreases, therefore resulting in an effectively uniform rewiring scheme.  In all cases, as expected~\cite{comin2014random}, the steering coefficient increases with the reciprocity.  However, this effect is particularly pronounced for the clustering-based rewiring.  This is happens because reducing the reciprocity between nodes with high clustering has an effect of diverting the random walk visits from these groups, reducing the steering coefficient. The degree-based rewiring resulted in a stronger effect on the steering than uniform alterations.

\begin{figure}[ht]
	\begin{center}
		\includegraphics[scale=0.50]{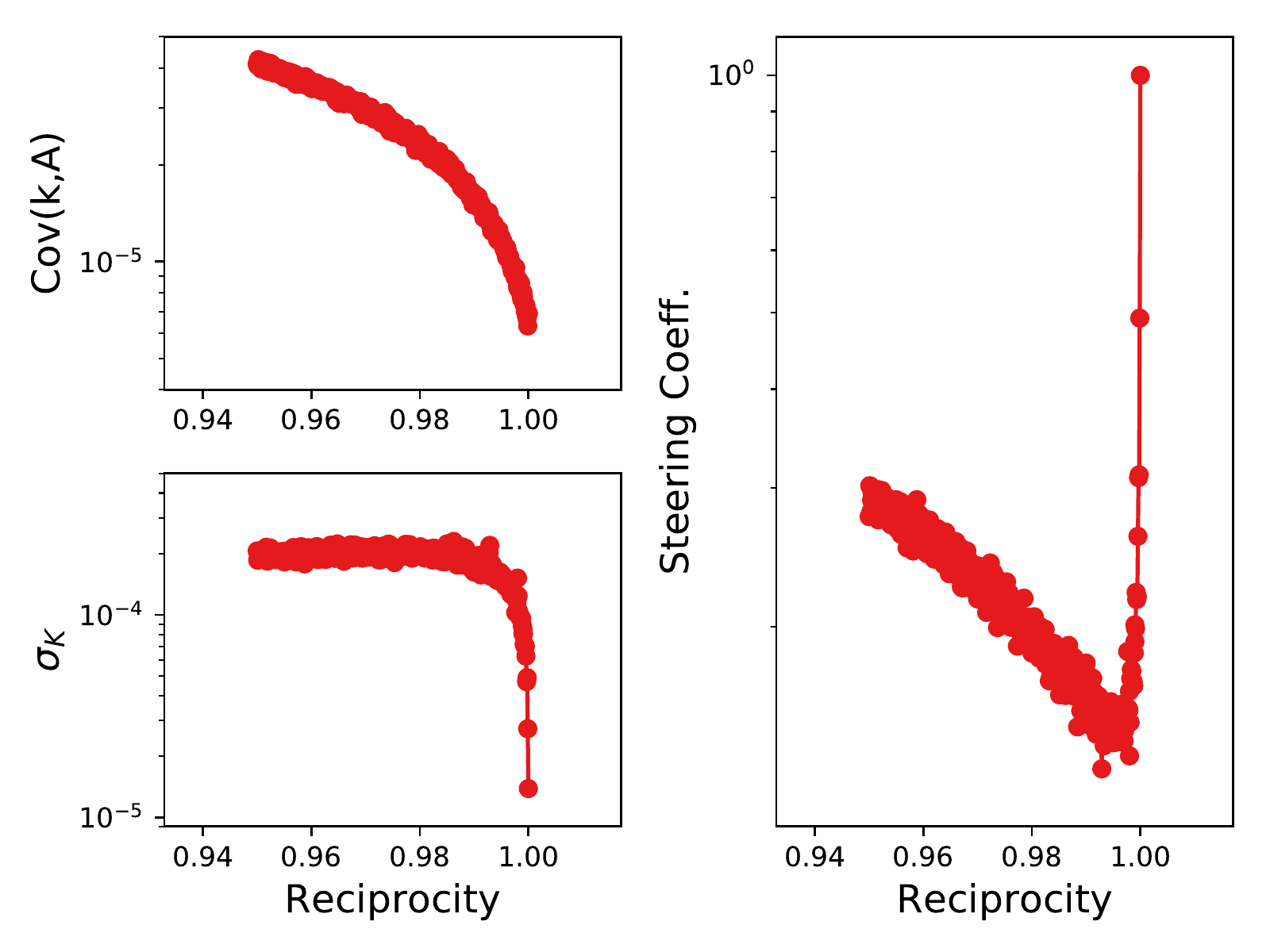}
    	\caption{Covariance between node degree and activation, standard deviation of node degree and the respective steering coefficient obtained inside the dashed rectangle shown in Figure~\ref{fig:theoretical} (for the WS model). A uniform rewiring strategy was used.}
		\label{fig:Wt_zoom}
	\end{center}
\end{figure}

\begin{figure}[ht]
	\begin{center}
		\includegraphics[scale=0.5]{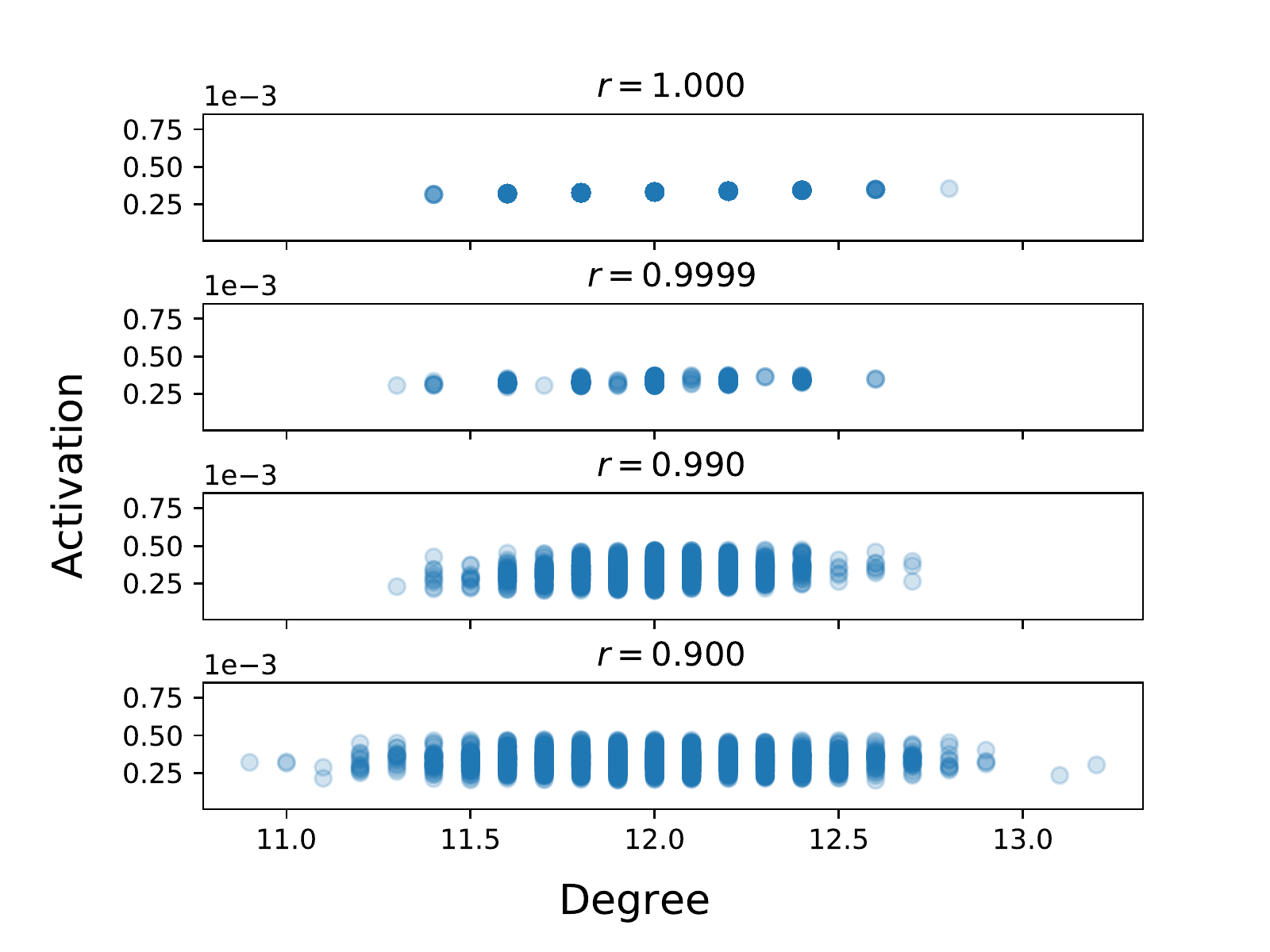}
    	\caption{Node activation versus degree for WS networks having distinct reciprocities, as indicated in the title of the plots.}
        \label{fig:ws_act_degree}
	\end{center}
\end{figure}

\begin{figure}[ht]
	\begin{center}
		\includegraphics[scale=0.7]{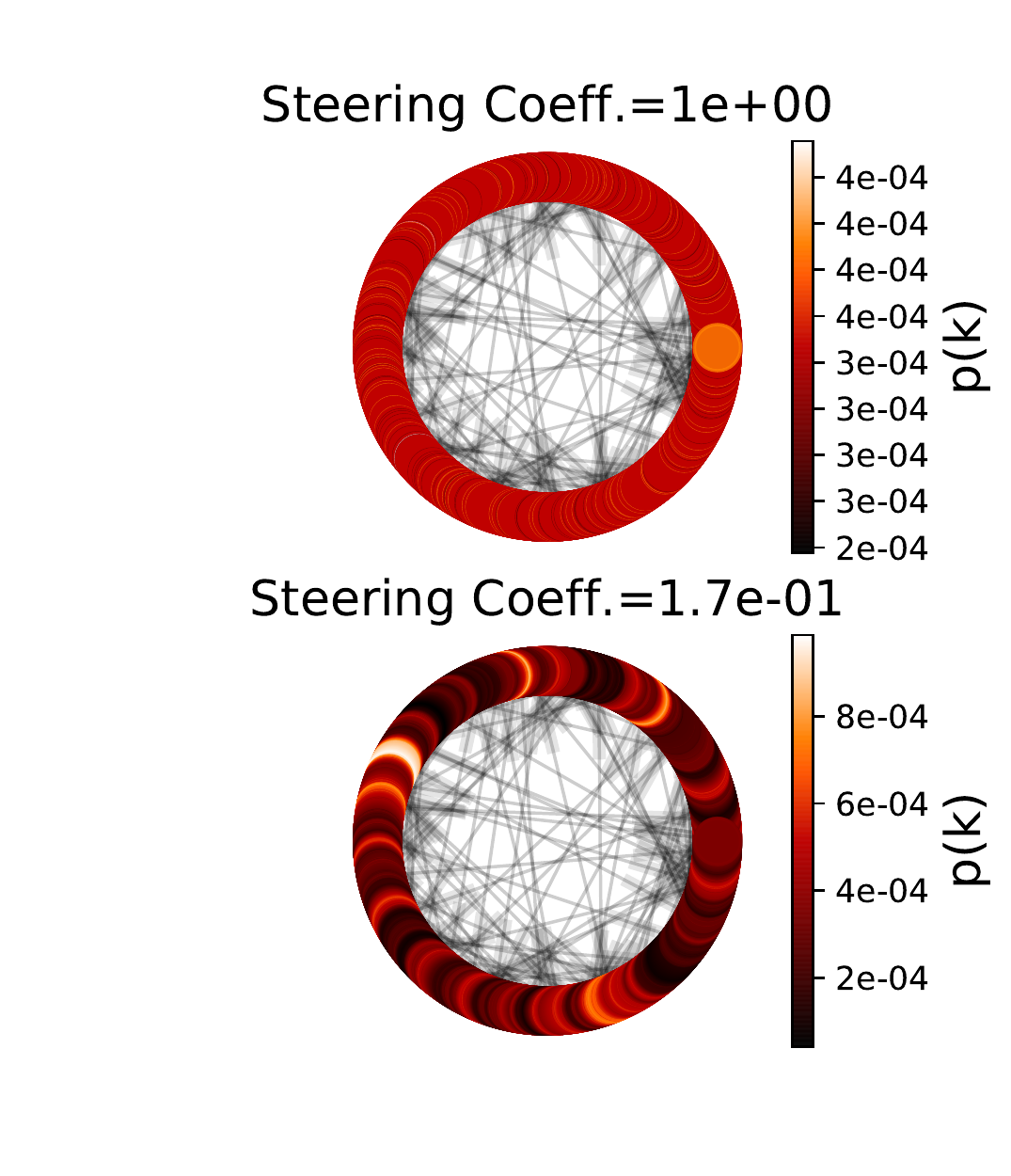}
    	\caption{A WS network having rewiring parameter $\rho=5\times 10^{-3}$ and reciprocity $r=1$ (upper plot) and $r=0.99$ (bottom plot). The activation of the nodes, $p(k)$, is represented by colors.}
		\label{fig:ws_color_map}
	\end{center}
\end{figure}

In the case of the WS topology, the effects of the four distinct rewiring methods are similar.  However, a remarkable effect occurs right before the maximum reciprocity is achieved.  More specifically, after decreasing steadily, the steering coefficient undergoes a steep increase reaching the maximum value (1.0).  In order to better understand this phenomenon, we separated the quantities that define the steering coefficient (itself corresponding to the Pearson correlation coefficient), and the results are shown in Figure~\ref{fig:Wt_zoom}.  This figure includes the covariance between node degree and activation (a), the standard deviation of the activation (b), and a zoom (c) focusing on the minimum of the curve shown in Figure~\ref{fig:theoretical} (indicated by a dashed square).  The Pearson correlation coefficient involves the division of the aforementioned covariance and the standard deviation of activation, and therefore reflects the abrupt variation of the latter at higher values of reciprocity.  In order to understand this abrupt variation of standard deviation, we performed additional simulations shown in Figures \ref{fig:ws_act_degree} and \ref{fig:ws_color_map}.   In Figure \ref{fig:ws_act_degree}, it is shown that the relationship between the degree and the activation is critically affected by small changes in the reciprocity.  Starting from the maximum reciprocity (1.0), the dispersion of the points in the space activation$\times$degree increases (for r = 0.9999 and 0.990) but then decreases (r=0.900).  This is the origin of the minimum in the curve in Figure \ref{fig:Wt_zoom}.  Figure~\ref{fig:ws_color_map} shows the values of the activation in a WS network for reciprocity near 0.990, where the minimun peak of the curve in Fig.~\ref{fig:ws_color_map} is observed.  At this reciprocity, the activation tends to concentrate in a few centers (yellow/white), implying in less relationship with the degree of the nodes.  

The Erd\H{o}s-R\'enyi model yielded  steering curves (Fig.~\ref{fig:theoretical}) which are, for higher values of reciprocity, almost indistinguishable.  Otherwise, slightly different steering coefficients are observed for the different types of rewiring.  Such similar steering coefficient values are a consequence of the fact that most nodes in ER networks have similar degree, clustering coefficient and matching indices. The obtained curves are close to being straight.   Similar results were obtained for the WX networks, with the difference that larger variations of the steering coefficient were observed for the degree, cluster and matching index-based rewirings. 

All in all, we have that the largest variations of the relationship between activation and degree, as quantified by the steering coefficient, take place for the WS networks.  This implies that reciprocity perturbations in these networks, such as induced by attacks, have a more substantial effect on the relationship topology$\times$dynamics.  Therefore, it becomes more difficult to predict the dynamics given the topology (degree) of these networks for larger values of reciprocity.  Contrariwise, the smallest influence of the reciprocity on the steering coefficient is observed for the ER networks, meaning that the relationship between topology and dynamics is less affected in this topology.

\section{Average degree comparison}

\begin{figure}[ht]
	\begin{center}
		\includegraphics[scale=0.43]{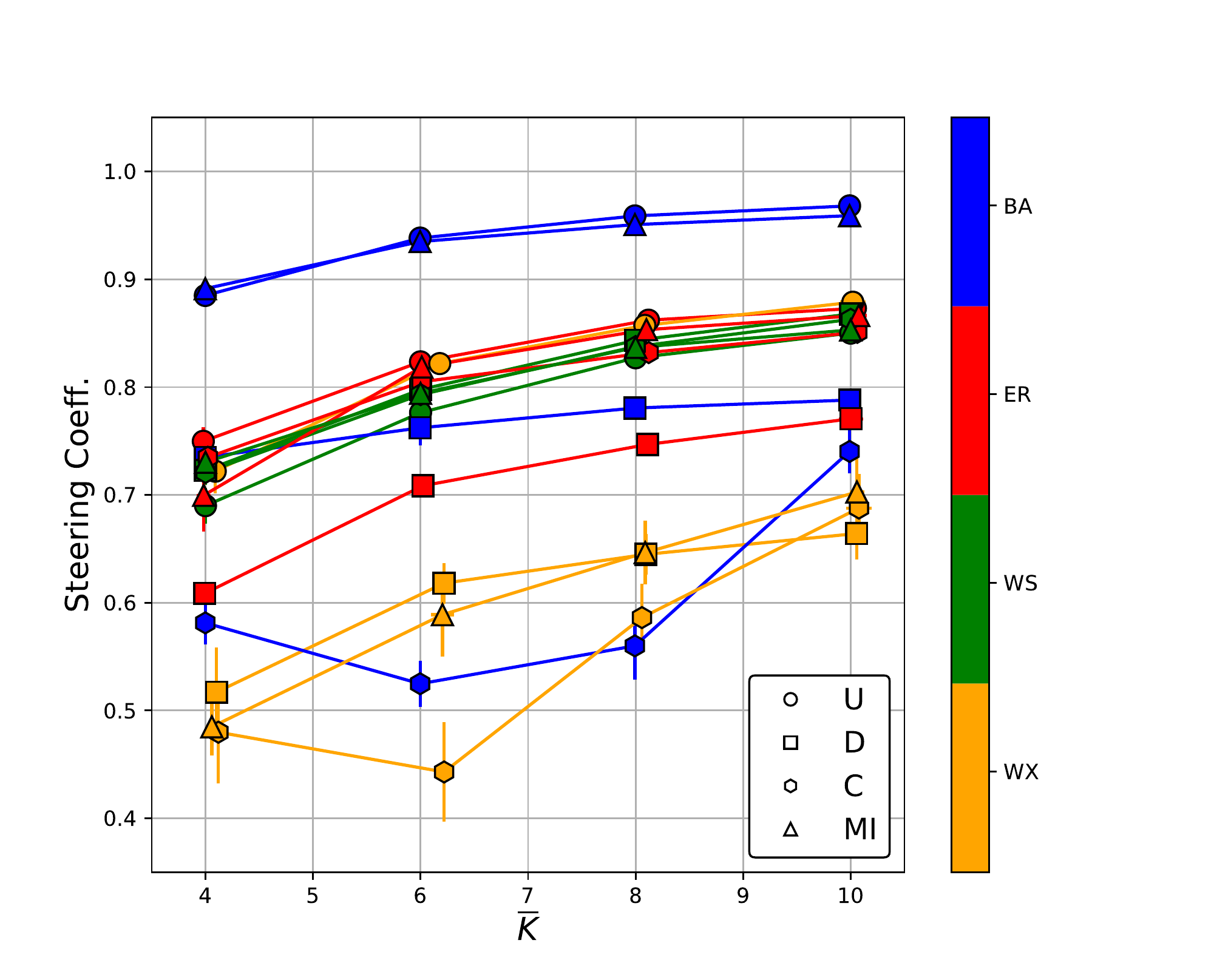}
		\caption{The steering coefficient in terms of average indegree, as obtained for the considered network models.  A reciprocity $r=0.5$ was adopted. The rewiring strategies are indicated by the markers, and the models by colors.}
    	\label{fig:Steer_X_AG}
	\end{center}
\end{figure}

Another relevant point addressed in this work concerns the relationship between the steering coefficient and the average degree of the networks.  Figure~\ref{fig:Steer_X_AG} shows that these two variables are correlated, for a fixed reciprocity $r=0.5$.

This result shows that, excluding the BA and the WX in the case of the clustering rewiring strategy, the steering coefficients are more influenced (in the sense of having larger derivative) for lower values of the average degree, defining a positive concavity curve.  In addition, larger steering coefficient values are obtained for larger node in-degrees.  This means that the activation can be better predicted from the degree in such situations.  We also observe that the clustering rewiring strategy for BA and WX presents a different curve, with a minimum located at $\bar{k}\approx 6$.

\subsection{Real-World networks}

\begin{figure}[ht]
	\begin{center}
    	\includegraphics[scale=0.55]{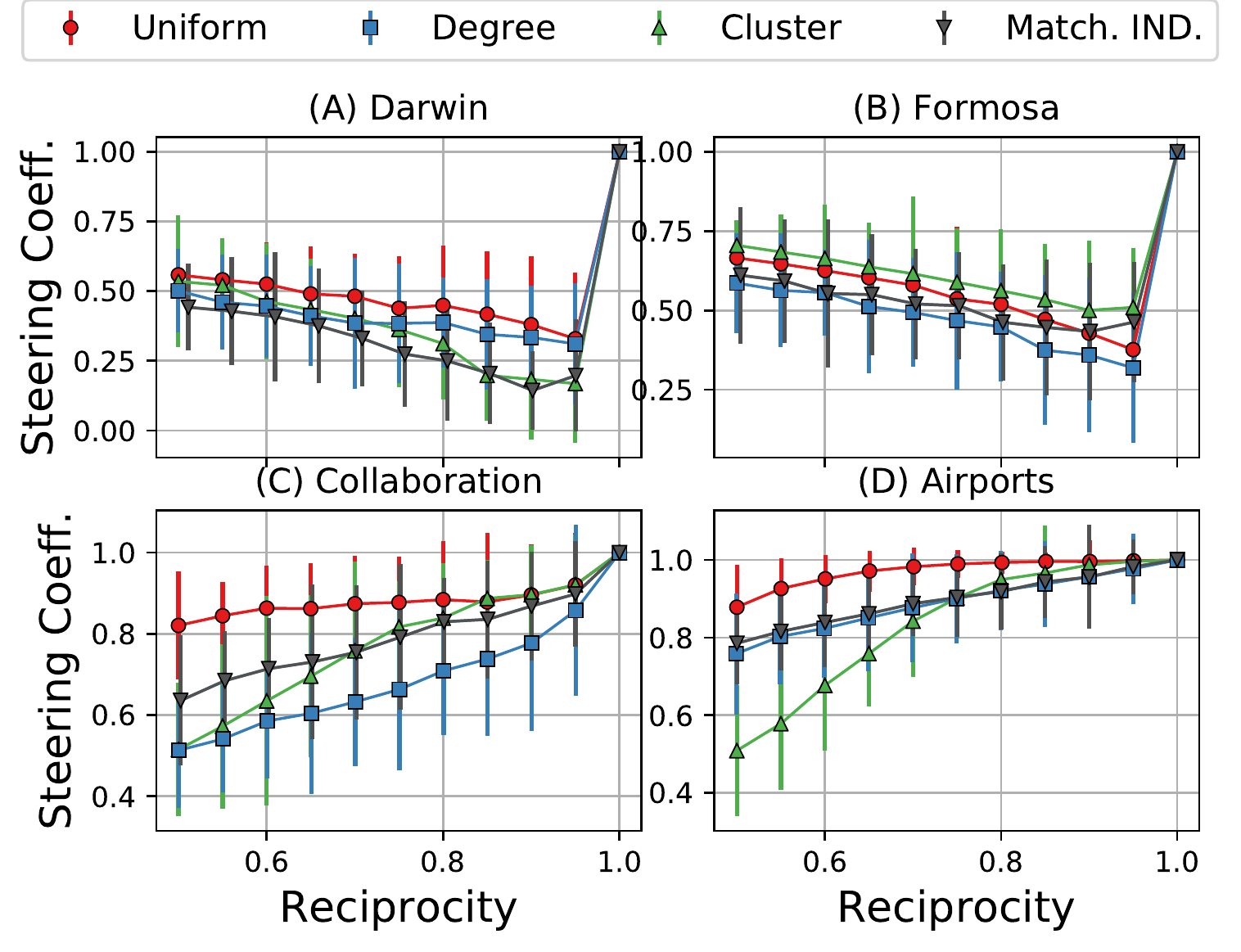}
    	\caption{Relationship between steering coefficient and network reciprocity for four real-world networks.}
		\label{fig:Real}
	\end{center}
\end{figure}

The investigation of the effect of several types of rewirings on the relationship between topology and activation was also performed with respect to four real-world networks, namely two street networks (Darwin and Formosa), airports and High Energy Physics collaborations.  These networks were chosen so as to have similar size (number of nodes) and average degree between themselves.  However, we observe slightly larger average degree in the airport network and slightly larger number of nodes for the collaboration network.  The characteristics of the chosen real-world networks are summarized in Table\ref{tab:properties}.  Figure~\ref{fig:Real} shows the relationship between steering coefficient and reciprocity for the real-world networks.

\begin{figure}[ht]
	\begin{center}
		\includegraphics[scale=0.5]{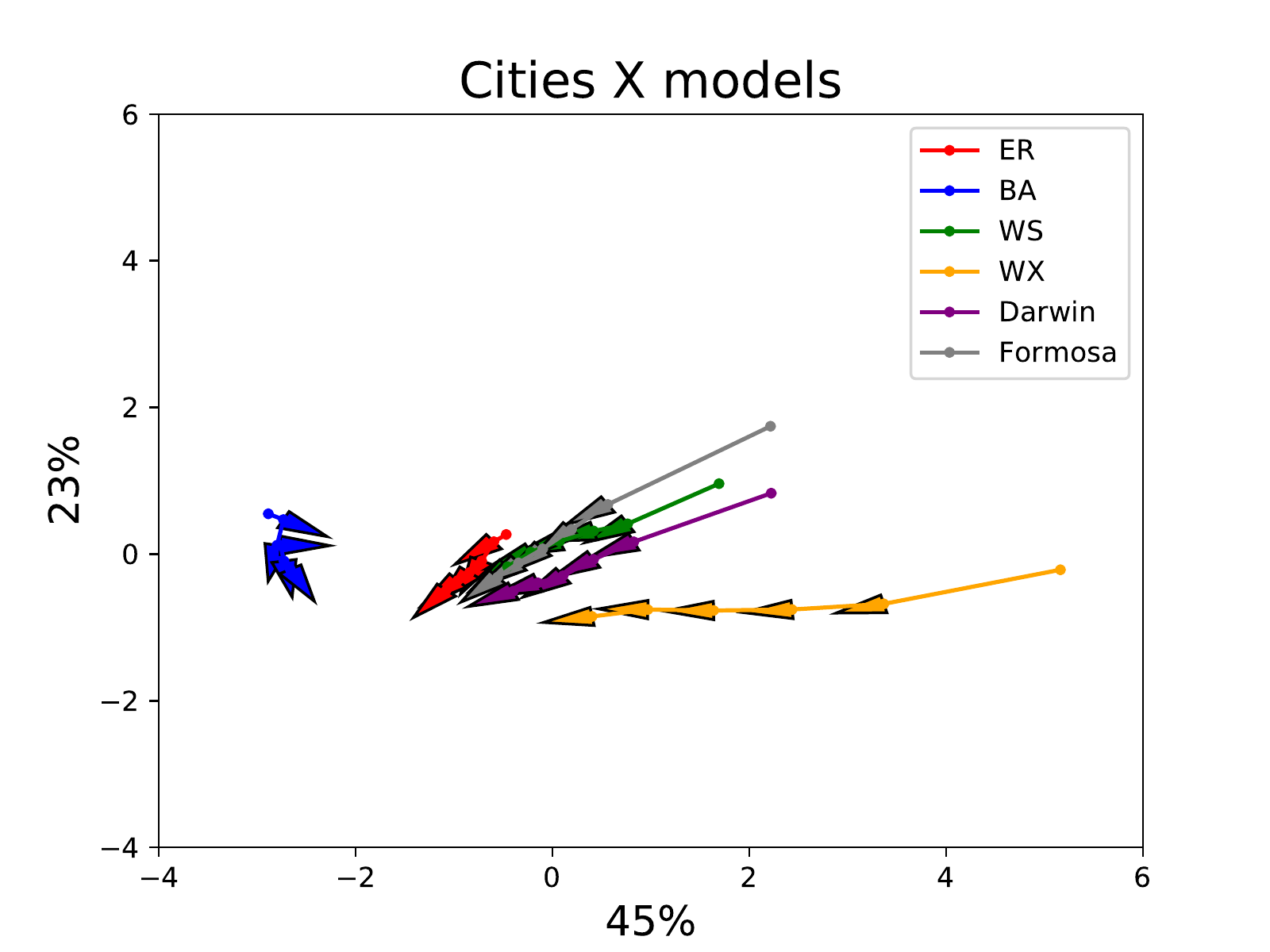}
		\caption{Principal Component Analysis (PCA) of the network models and the two street networks, Darwin and Formosa.}
		\label{fig:PCA_cities}
	\end{center}
\end{figure}

A small difference between the variation of the steering coefficient in terms of the reciprocity for the four considered rewiring schemes can be observed (Figures~\ref{fig:Real}(A) and (B)) for both Darwin and Formosa.  In addition, similar results were obtained between these two cities, both closely resembling the effects observed for the WS model (as discussed in the previous section).  This suggests that the topology of these two cities are well-modeled by WS networks.  In order to check this possibility, we performed a systematic comparison between the topology of these two cities, as expressed by several measurements (average matching index, average clustering coefficient, average shortest path length, average accessibility for level 3~\cite{viana2012effective}, standard deviation of degree, and assortativity), and all the four considered theoretical network models.  All used networks have, as much as possible, the most similar number of nodes and average degree.   In order to have a more comprehensive characterization, we performed progressive uniform rewiring of the link directionality for all considered networks.   The results are shown in Figure~\ref{fig:PCA_cities} in terms of the principal component analysis (PCA).  The PCA projection allowed preservation of 68\% of the variance in its two first axes, substantiating the relevance of this projection.  The original and rewired instances of the Darwin and Formosa cities resulted near the WS networks, confirming the initial hypothesis that these two cities have topology substantially similar to WS.

\begin{figure}[ht]
	\begin{center}
    	\includegraphics[scale=0.5]{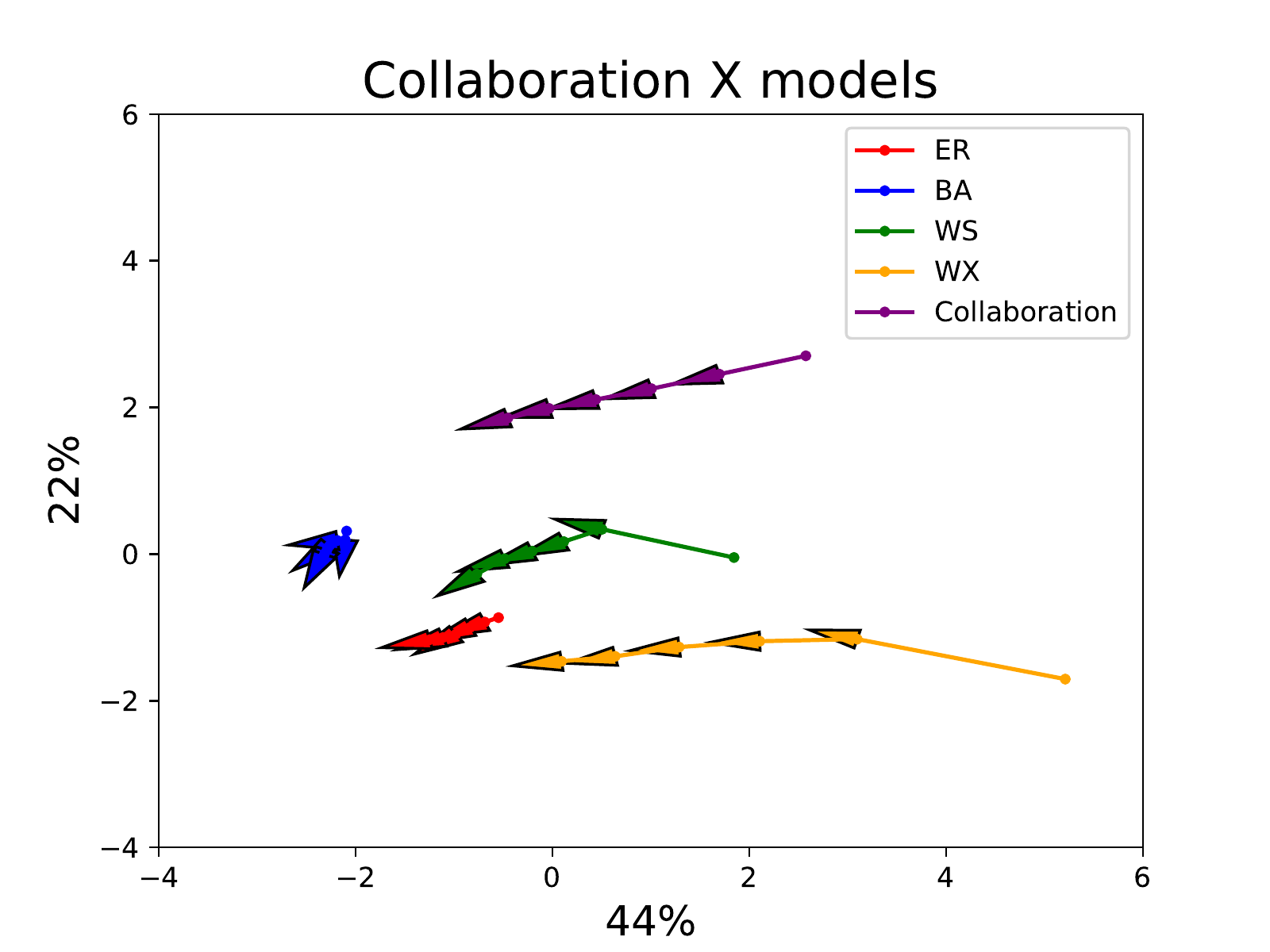}
    	\caption{PCA of the network models and the Collaboration network.}
    	\label{fig:PCA_hep}
	\end{center}
\end{figure}

Figure~\ref{fig:Real}(C) presents the steering coefficient in terms of the reciprocity obtained for the Collaboration network.  The uniform rewiring strategy yielded the smallest variation of the steering coefficient.  Contrariwise to what was found for the two cities, the results obtained for the Collaboration network do not closely resemble any of the curves obtained for the four theoretical network models.  This suggests that the topology of the Collaboration network departs significantly from those exhibited by all the considered models.  This has been indeed confirmed by the PCA comparison (performed analogously to the cities analysis), shown in Figure~\ref{fig:PCA_hep}. 

\begin{figure}[ht]
	\begin{center}
    	\includegraphics[scale=0.5]{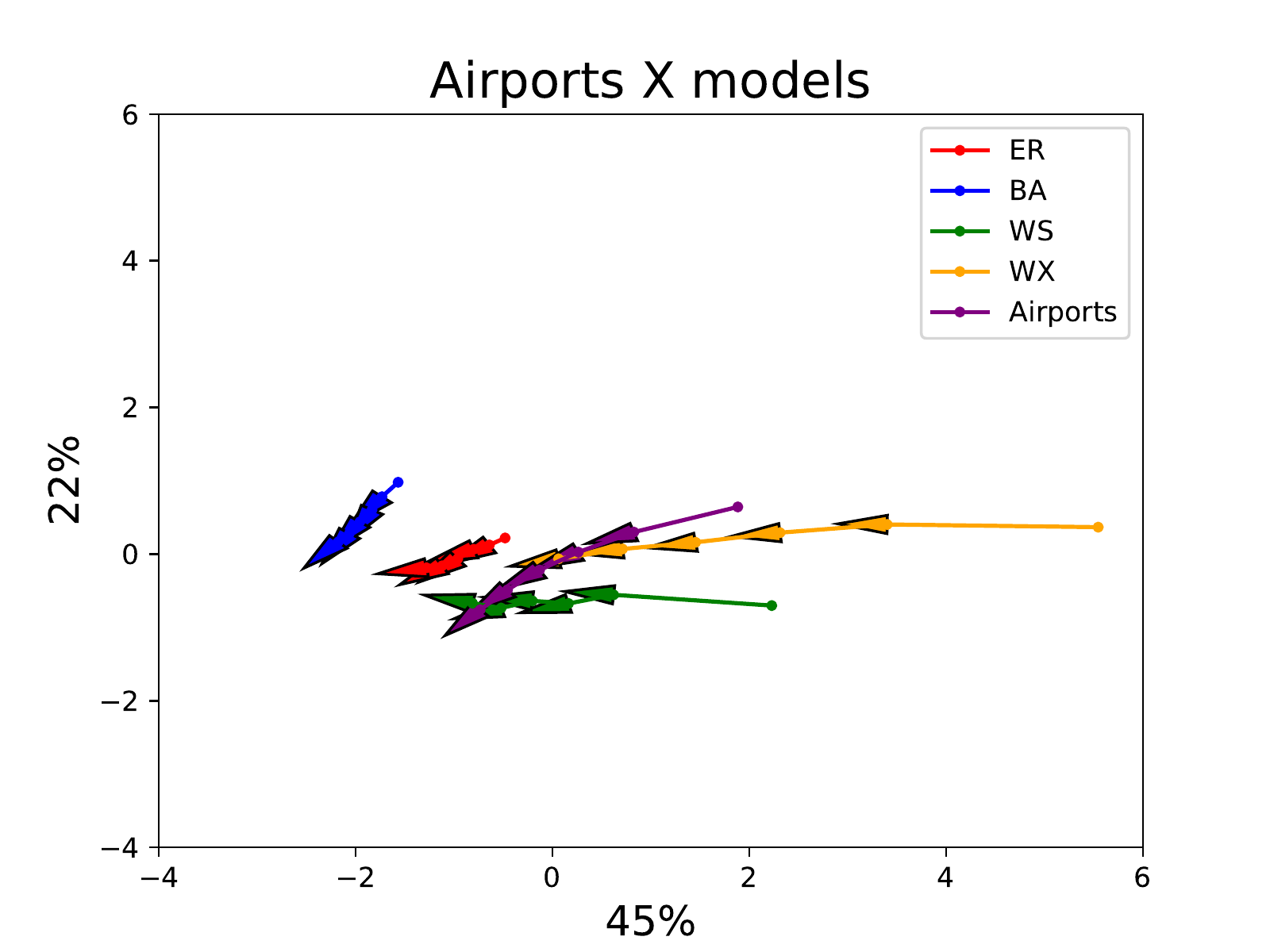}
    	\caption{PCA of the network models and the airport network.}
    	\label{fig:PCA_air}
	\end{center}
\end{figure}

The study of the Airports network is depicted in Figure~\ref{fig:Real}(D). The smallest variation of the steering coefficient was again obtained for the uniform rewiring.  Remarkably similar variation of the steering coefficient in terms of the reciprocity was observed for the rewiring according to matching index and degree.  A markedly steep increase of steering coefficient was obtained with the clustering coefficient rewiring.  As with the Collaboration network, no clear relationship can be established between the observed curves and those obtained for the four theoretical models, which is confirmed by the PCA results shown in Figure~\ref{fig:PCA_air}.

% \begin{table}[ht]
% 	\begin{tabular}{|l*{6}{c}r}
% 		PCA & $\overline{\mu}$ & $\overline{cc}$ & $\sigma_K$ & $\overline{l}$ & assor. & acce. \\
% 		\hline	
% 		Col. & 0.53 & 0.5 & 0.27 & 0.44 & 0.12 & 0.34\\
% 		Cities & 0.46 & 0.49 & 0.46 & 0.40 &  0.10 & 0.40  \\
% 		Airp. & 0.54 & 0.52 & 0.13 & 0.50  & 0.09  & 0.31  
%         \end{tabular}
%         \caption{PCA absolute eigenvalues for x axis}
%         \label{table:PCA_x}
% \end{table}

% \begin{table}[ht]
% 	\begin{tabular}{|l*{6}{c}r}
% 		PCA & $\overline{\mu}$ & $\overline{cc}$ & $\sigma_K$ & $\overline{l}$ & assor. & acce. \\
% 		\hline	
% 		Col. & 0.11 & 0.25 & 0.64 & 0.23  & 0.65 & 0.08 \\
% 		Cit. & 0.42  & 0.41 & 0.31 & 0.08 & 0.54 & 0.47  \\
% 		Airp. & 0.11 & 0.19 & 0.06  & 0.20 & 0.78 & 0.49  
%         \end{tabular}
%         \caption{PCA absolute eigenvalues for y axis}
%         \label{table:PCA_y}
% \end{table}

\section{Conclusion} 

Much of the current research in network science focus on the key issue of how topology and dynamics can be related in complex networks.  Despite the importance of this subject, much still remains to be discovered and characterized regarding this relationship with respect to varying network models and types of dynamics.  One interesting possibility consists in using the steering coefficient as an indicator of this relationship, as it provides a direct indication of how the dynamics can be predicted from the topology of a given network.  As shown in recent works~\cite{comin2014random,de2017network}, directed networks imply more intricate relationships between topology and dynamics.   In the present work, we focus this type of networks, using different types of rewiring strategies (based on node degree, node clustering coefficient, edge matching index coefficient, as well as a uniform approach independent of the network topology) and assessing their effect on the respective steering coefficients.

Indeed, the main obtained results show that reciprocity plays a particularly important role in determining the steering coefficient and, hence, the relationship between topology and dynamics. More specifically, we observed that this relationship is affected in different ways depending on the applied rewiring method.  Computational simulations have shown that the different types of rewirings can have strong effect on the steering coefficient. In most models, rewiring according to the clustering coefficient presented the greatest influence on the steering coefficient, followed by the degree and matching index. 

The changes in steering coefficient observed for varying reciprocity were also evaluated for different models. Steering coefficient values increasing with the reciprocity were observed for the BA, WX and ER models, being more linear in the latter two cases.  A decreasing behavior, followed by a steep increase, was observed for WS networks.  A possible explanation for this effect was proposed, based on large standard deviations observed for reciprocities near one.  For the real-world networks, a great similarity of steering coefficient behavior was observed between the cities and the WS model. For the other considered real-world networks, namely airport and collaboration the PCA approach showed that those networks are relatively distant from all the considered network models. 

In order to investigate the effect of the average in-degree on the steering coefficient, we performed rewirings in the theoretical networks while varying the average degree, while keeping the reciprocity fixed. The results show that the average degree are, with few exceptions, positively related to the steering coefficient. 

In addition to their theoretical contributions and implications, the obtained results can be considered in several practical situations.  For instance, they provide guidelines for designing or modifying networks in order to obtain a steering coefficient that remains more stable under the effect of perturbations.  In other words, the degree in which the topology influences the dynamics in a complex system can be projected and/or controlled by using rewiring schemes that affect the steering coefficients in the ways reported in this work.  In principle, the most suitable degree of relationship will depend on each specific situation.  For instance, if topological changes are frequent in a system, one may be interested in having the respective dynamics to follow less directly the topology, implying in smaller steering coefficients.  This could be the situation in transportation or communication systems.  Contrariwise, if the objective is that the dynamics will reflect even small topological changes, such as in monitoring or measurement systems, a larger steering coefficient would be sought.

Several future investigations related to the topics developed in this work can be considered, such as using other types of dynamics, topologies, and rewirings.

\acknowledgments
C. H. Comin thanks FAPESP (grant no. 15/18942-8) for financial support. L. da F. Costa thanks CNPq (grant no. 307333/2013-2) and NAP-PRP-USP for sponsorship. This work has been supported also by FAPESP grants 11/50761-2 and 2015/22308-2.

\bibliographystyle{apsrev}
\bibliography{references}

\end{document}